\documentclass[onecollarge,natbib]{svjour2}
\bibpunct{[}{]}{,}{n}{}{,} 
\smartqed  
\usepackage{graphicx}
\journalname{Few-Body Systems (APFB2011)}
\begin{document}

\title{\boldmath
$CP$ violation and strong pion-pion interactions in the weak  $B^\pm \to \pi^\pm \pi^\mp \pi^\pm$ decays
\thanks{Presented at the fifth 
    Asia-Pacific Conference on Few-Body Problems in Physics,
  Seoul, Republic of Korea, August 21-26, 2011. This work was partially supported by the IN2P3-Polish Laboratories Conventions (project No 08-127).
}}


\author{B.~Loiseau        \and
        J.-P.~Dedonder \and
        A.~Furman \and
        R.~Kami\'nski \and
        L.~Le\'sniak 
}


\institute{B.~Loiseau \and J.-P.~Dedonder \at
             Laboratoire de Physique Nucl\'eaire et de Hautes \'Energies, Groupe Th\'eorie, 
Universit\'e Pierre et Marie Curie et Universit\'e Paris-Diderot, IN2P3 et CNRS, 4 place Jussieu, 75252 Paris, France  \\
              \email{loiseau@lpnhe.in2p3.fr}           
           \and
          A.~Furman   \at
          ul. Bronowicka 85/26, 30-091 Krak\'ow, Poland
          \and
            R. Kami\'nski \and L. Le\'sniak  \at
         Division of Theoretical Physics, The Henryk Niewodnicza\'nski Institute of Nuclear Physics,
                  Polish Academy of Sciences, 31-342 Krak\'ow, Poland    
}

\date{Received: date / Accepted: date}

\maketitle

\begin{abstract}
An analysis of $B^\pm \to \pi^\pm \pi^\mp \pi^\pm$ decay data in a quasi two-body 
$B^\pm \to \pi^\pm [\pi^+ \pi^-]_{S,P,D}$  QCD factorization framework is performed.
The short distance amplitudes are calculated including next-to-leading order in $\alpha_s$ vertex and penguin corrections.
The long distance amplitudes due to the $ [\pi^+ \pi^-]_{S,P,D}$ final state interactions are described by the pion non-strange scalar and vector form factors for the  $[\pi^+ \pi^-]_{S}$ $S$-wave state and $[\pi^+ \pi^-]_{P}$ $P$-wave state, respectively and by a relativistic Breit-Wigner formula for the $[\pi^+ \pi^-]_{ D}$ $D$-wave state.
We achieve a good fit of the data with only three parameters for the $S$ wave, one for the $P$ wave and one for the $D$ wave.
Our model gives a unified unitary description of the three scalar resonances, $f_0(600)$, $f_0(980)$ and $f_0(1400)$ in terms of the pion scalar form factor. We predict for the $B$ to $f_2(1270)$ transition form factor, $F^{Bf_2}(m_\pi^2)=0.098\pm0.007$.
\end{abstract}

\section{Introduction}
\label{intro}

\paragraph{Motivations} 
Three-body hadronic charged $B$ decays into three charged pions are interesting
since they allow  i) to study $CP$ violation, ii) to test the standard model and QCD.
The weak and strong phase differences are a measure of the $CP$ asymmetry $A_{CP}$.
It can be obtained from the difference between the $B^+\to \pi^+ \pi^- \pi^+$ and $B^-\to \pi^- \pi^- \pi^+$  branching fractions.
For instance the recent BABAR Collaboration isobar model Dalitz-plot analysis~\cite{Aubert:2009} gives for the $B^\pm$ decay into $\rho(770)^0 \pi^\pm$,

\begin{equation}
\label{ACP}
A_{CP}[B^\pm \to \rho(770)^0\pi^{\pm},\ \rho(770)^0\rightarrow \pi^+\pi^-]=\left(18\pm7\pm5^{+2}_{-14}\right)\%.
\end{equation}
These rare weak decays arise from $W$-boson exchange, which, together with the fact that the $b$ quark mass, $m_b$, is large leads, for the short distance amplitude, to a systematic perturbative calculation in a QCD framework.
One further assumes factorization which amounts to express the long distance amplitude in terms of form factors~\cite{NPB170}.
The final state hadronic interactions introduce however important sources of uncertainties.
 


\paragraph{QCD factorization for quasi two-body decays}
The $B$ meson decays into two mesons $M_1$ and $M_2$ are well described~\cite{NPB675} within the QCD factorization scheme.
 If no QCD factorization formalism exists for three-body $B$ decays, one can however use a quasi two-body approach.
In the $B$ rest frame of the three-body $B^\pm \to \pi^\pm \pi^- \pi^+$ decays and  for low invariant $\pi^- \pi^+$  mass, $m_{\pi^- \pi^+} < 2$~GeV, the $\pi^+$ and $\pi^-$ mesons of the  $ \pi^- \pi^+$ pair move more or less in the same direction.
On can then reduce the study of the  three-body $B^\pm \to \pi^\pm \pi^- \pi^+$ decays to that of  quasi two-body $B^\pm \to \pi^\pm [\pi^- \pi^+]$ decays provided we make the hypothesis that the states of the $ [\pi^- \pi^+]$ pair originate from a quark-antiquark $q\bar q$ state.

Such an approach has been recently applied by some of us to  $B \to K \pi^+\pi^-$ decays by studying $B \to K[\pi^+\pi^-]_{S,P}$~\cite{fkll,PRD74} and  $B \to \pi^\pm [K\pi^\mp]_{S,P}$~\cite{PRD79,Leitner_PRD81}.
In these works,  the long distance $[\pi^+\pi^-]_{S}$ $S$-wave state includes the scalar $ f_0(600)$ and  $ f_0(980)$ resonances~\cite{fkll,PRD74} while the $P$-wave $[\pi^+\pi^-]_{P}$ state~\cite{PRD74} takes into account the vector meson $\rho(770)^0$.
In Refs.~\cite{PRD79,Leitner_PRD81} the long range $S$-wave $ [K\pi^\mp]_{S}$ state incorporates the scalar $\kappa(800)$, $K^*_0(1430)$ resonances and the $P$-wave $[K\pi^\mp]_{P}$ pair the vector meson $K^*(892)$.

In the present work we introduce the quasi two-body decays $B^\pm \to \pi^\pm [\pi^+ \pi^-]_{S,P,D}$ including, for the long range amplitude in $ [\pi^+ \pi^-]_{S}$ the scalar $f_0(600), f_0(980), f_0(1400)$, in  $[\pi^+ \pi^-]_{P}$, the vector $\rho(770)^0, \rho(1450)^0,  \rho(1700)^0$ and in  $[\pi^+ \pi^-]_{D}$ the tensor $f_2(1270)$.
For the short distance amplitudes we shall introduce the leading order (LO) and the next-to-leading order (NLO) vertex and penguin corrections.
A full account of the work we present here can be found in Ref.~\cite{ActaB42}.

 \begin{figure}
\centering
  \includegraphics[angle=0,width=0.43\columnwidth]{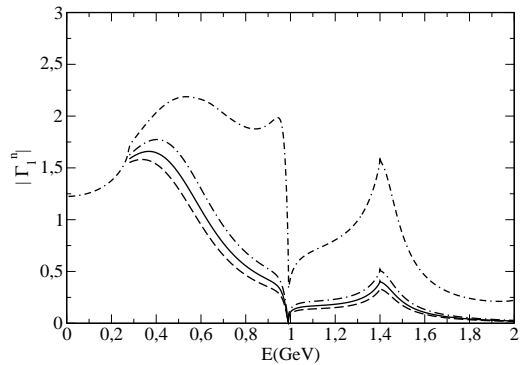}
 \caption{Solid line: modulus of the pion scalar form factor  $\Gamma_1^{n*}$ from our fit using the NLO effective Wilson coefficients. The parameter $\kappa$, ensuring the convergence of our integral equation~\cite{ActaB42}, is  $2$~GeV.  The fitted parameter $c$,  controlling the high energy behavior, is $c=(19.5 \pm 4.2)$~GeV$^{-4}$. Double-dash dot line: modulus of the pion scalar form factor calculated in Ref.~\cite{L6} from the Muskhelishvili-Omn\`es equations. The dash-dot line and the dashed one represent the variation of the  $\Gamma_1^n$ modulus when $c$ varies within its error band.
 }
 \label{fig:1}       
 \end{figure}

\section{Decay amplitudes}
\label{QCDFamp}

\paragraph{QCD factorization for $B^\pm \to \pi^\pm [\pi^+ \pi^-]_{S,P,D}$ }
In the QCD factorization approach one performs  a perturbative expansion of decay amplitudes  in terms of two small parameters, $\Lambda_{QCD}/m_b$  and $\alpha_s$, the strong coupling constant.
Using operator product expansion, one writes

\begin{equation}
\label{Heffscheme}
\langle  \pi^\pm [\pi^+ \pi^-]_{S,P,D}\vert H_{\textrm{eff}}\vert B^\pm\rangle=\frac{G_F}{\sqrt{2}}
V_{CKM} \sum_i C_i(\mu) \langle  \pi^\pm [\pi^+ \pi^-]_{S,P,D}\vert O_i(\mu)\vert B^\pm\rangle,
\end{equation}
where $G_F=1.166\times 10^{-5}$ GeV$^{-2}$ is the Fermi coupling constant and $H_{\textrm{eff}}$ is the effective weak Hamiltonian given in Eq.~(1) of Ref.~\cite{ActaB42}. 
The $V_{CKM}$ factor arises from product of  Cabibbo-Kobayashi-Maskawa matrix elements giving the strength of the coupling of the quarks to the $W$ boson.
The $O_i(\mu)$ are local operators and the $C_i(\mu)$ are the short range Wilson coefficients coming from the $W$  exchange and calculated at the renormalization scale $\mu$,  $\mu \sim  m_b$.
Furthermore, schematically, the ansatz of factorization reads,

\begin{equation}
\label{M1M2factorisation}
\langle M_1M_2\vert O_i(\mu)\vert B\rangle = \langle M_1\vert j_1\vert 0\rangle
 \langle M_2\vert j_2\vert B\rangle
 \left[1+\Sigma_{n=1}^\infty
 r_n\alpha_s^n+O(\Lambda_{QCD}/m_b) \right],
\end{equation}
where $j_{1,2}$ are bilinear quark currents.
The matrix element $\langle M_1\vert j_1\vert 0\rangle$ is given in terms of the decay constant of the meson $M_1$ while $\langle M_2\vert j_2\vert B\rangle$ can be calculated using the $B$ to $M_2$ transition form factor.
At the weak vertices, $r_n$ represent radiative corrections such as vertex or hard gluon exchanges with the spectator quark contributions~\cite{NPB675}.
Here the vertex corrections are taken into account but not the divergent
hard gluon or the annihilations corrections.

%
\begin{figure*}
 \centering
  \includegraphics[height=.25\textheight]{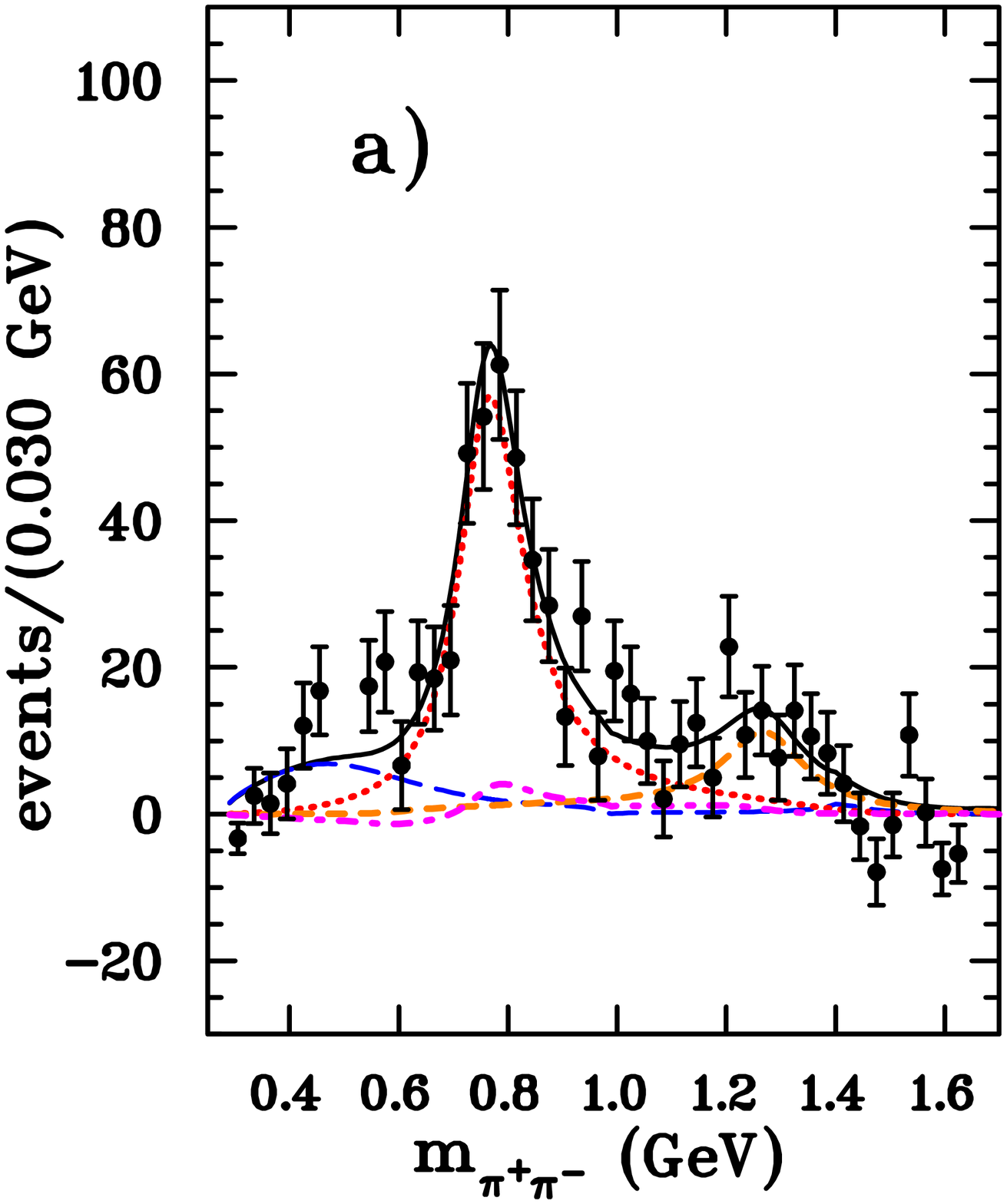}~~~~
   \includegraphics[height=.25\textheight]{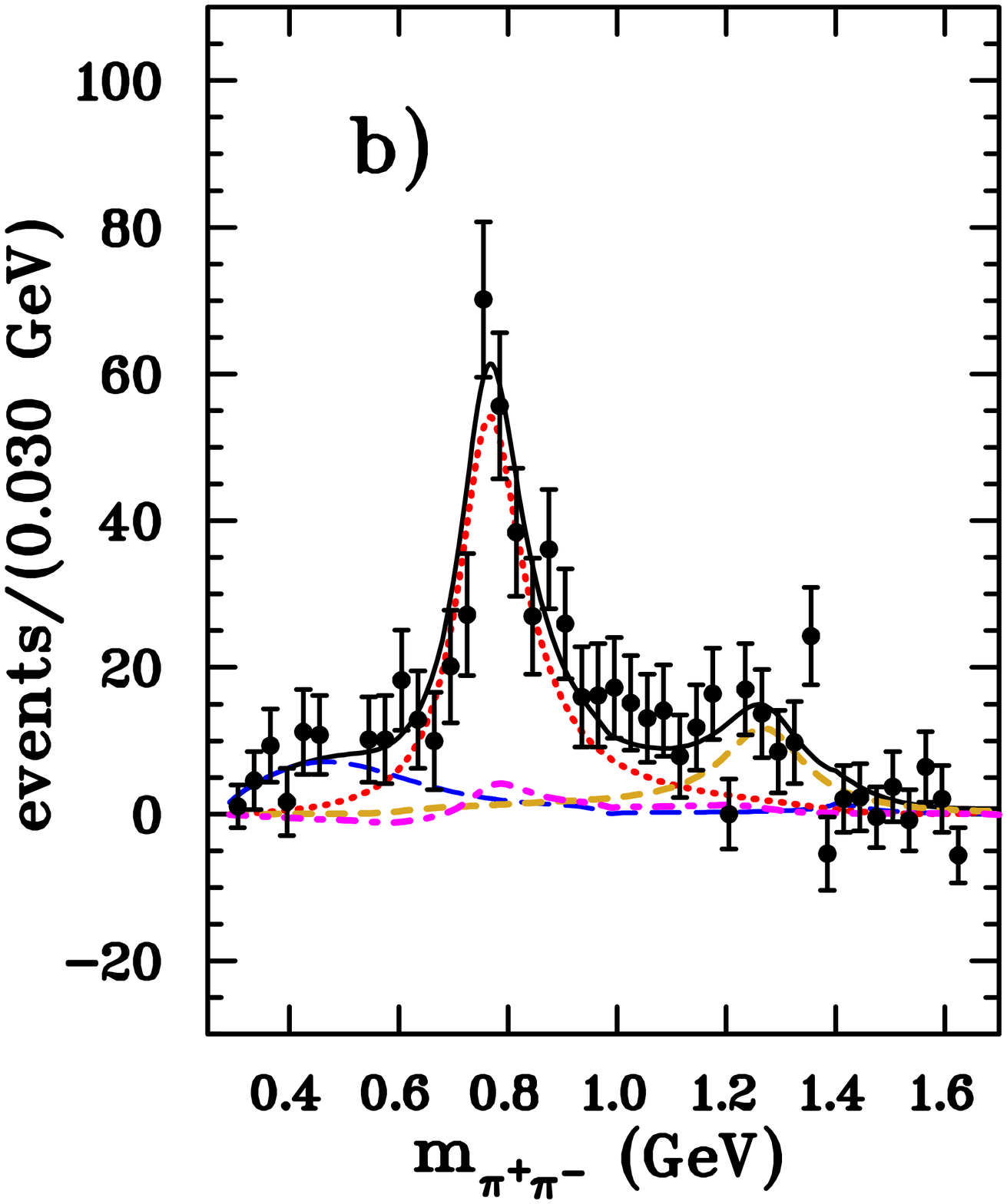}~~
 \caption{The  light effective $m_{\pi^+\pi^-}$ distributions from the fit to the BABAR data~\cite{Aubert:2009}, a) for the $B^-$  and b) for the $B^+$ decays. 
The long-dash line corresponds to the $S$-wave, the dot line 
to the $P$ wave, the short-dash line to the $D$ wave and the dot-dash line to the interference  contributions. 
The solid line corresponds to the sum.}
 \label{fig:2}       
 \end{figure*}

The factorization of the matrix element
$\left \langle  \pi^\pm(p_1) [\pi^+(p_2){\pi^-}(p_3) ]_{{S},{P},D}\  \vert j_1\otimes j_2\vert B^\pm(p_B) \right \rangle$
gives two terms.
The first term leads to the  long distance functions:

 \begin{equation}
 \label{SP}
 \left \langle \pi^\pm(p_1)\vert j_1\vert B^\pm(p_B)\right \rangle \left \langle  [\pi^+(p_2){\pi^-}(p_3) ]_{S,P}\vert j_2\vert 0\right \rangle
\propto \ F_{0,1}^{B\pi}(q^2)  F_{0,1}^{\pi \pi}(q^2), 
\end{equation}
where $q=p_B-p_1=p_2+p_3$.
The $F_{0,1}^{B\pi}(q^2)$ are the scalar or vector $B$ to $\pi$ transition form factor taken from light cone sum rules~\cite{PRD71} and the
$ F_{0,1}^{\pi \pi}(q^2)$  the scalar or vector pion form factor described below.
The second term leads to the long distance terms:

  \begin{equation}
 \label{SPD}
 \left \langle  [\pi^+(p_2){\pi^-}(p_3) ]_{S,P,D} \vert j_1\vert B^\pm(p_B)\right \rangle  \left \langle \pi^\pm(p_1)\vert  j_2 \vert 0 \right \rangle 
\propto G_{R_{S,P,D} \pi^+\pi^-}^n(q^2) \left \langle  R_{S,P,D} \vert j_1\vert B^\pm(p_B)\right \rangle \left \langle \pi^\pm \vert  j_2\vert 0 \right \rangle. 
 \end{equation}
The vertex functions  $G_{R_{S,P,D} \pi^+\pi^-}^n(q^2)$ describe the resonance $R_{S,P,D}$ decays into $\pi^+ \pi^-$ pairs.
The matrix elements  $\left \langle  R_{S,P,D} \vert j_1\vert B^\pm \right \rangle$ correspond to the $B^\pm$ to $R_{S,P,D}$ transition form factors for which we take $R_S \equiv f_0(980)$, $R_P\equiv \rho(770)^0$ and $R_D \equiv f_2(1270)$.
The term $ \left \langle \pi^\pm \vert  j_2\vert 0 \right \rangle$ is related to the pion decay constant $f_\pi$.
In our model, we assume 
 the  $G_{R_{S,P} \pi^+\pi^-}^n(q^2), G_{R_{D} \pi^+\pi^-}^n(q^2)$ vertex functions to be proportional to the
scalar or vector pion form factor  and to a relativistic Breit-Wigner amplitude~\cite{Aubert:2009} for the resonance $f_2(1270)$, respectively.
From the matrix elements given in Eqs.~(\ref{SP}) and (\ref{SPD}) one obtains the total 
symmetrized amplitudes for $B^\pm \to \pi^\pm \pi^\mp \pi^\pm$ decays as given in Eqs.~(21) to (26) of Ref.~\cite{ActaB42}.

\paragraph{Scalar form factor} The normalized pion scalar form factor, $\Gamma_1^{n*}(q^2)$, which is proportional to the form factor $F_0(q^2)= \sqrt{3/8} \left\langle [\pi^+\pi^-]_{{S}} \ \vert u \bar u+d \bar d \vert 0  \right \rangle $,
 is calculated from a unitary relativistic coupled channel model including $\pi \pi, K \bar K$ and effective $(2\pi) (2\pi)$ interactions (see Eqs.~(27) and (28) of Ref.~\cite{ActaB42}).
 The needed  $\pi \pi$ scattering  $T$ matrix corresponds to the solution $A$ of  Ref.~\cite{EPJC9}.
 The result of our fit (see next section) for the modulus of $\Gamma_1^{n*}(q^2)$ is plotted in Fig.~\ref{fig:1}.

 \paragraph{Vector form factor} The pion vector form factor results from the Belle Collaboration analysis~\cite{Fujikawa_PRD78_072006} of high statistics  $\tau^- \to \pi^- \pi^0 \nu_\tau$ decay data with a Gounaris-Sakurai model for the  three vector resonances $\rho(770)$, $\rho(1450)$ and $\rho(1700)$.
We use their parameters given in the third column of their Table VII.

\section{Fit to $B^\pm \to \pi^\pm \pi^\mp \pi^\pm$ decay data  }
\label{Fitto data}

We fit  the $B^\pm \to \pi^\pm \pi^\mp \pi^\pm$ BABAR Collaboration data with three parameters for the $S$ wave, one for the $P$ wave and one, $F^{Bf_2}(m_\pi^2)$, for the $D$ wave.
The $\chi^2$ fit to  the 170 data points from BABAR's invariant mass 
distributions together with the experimental branching ratio  $\mathcal {B}_P$ for 
$B^\pm \to \rho(770)^0 \pi^\pm,\  \rho(770)^0 \to \pi^+ \pi^- =(8.1\pm0.5\pm1.2^{+0.4}_{-1.1})\times 10^{-6} $ 
is  $\chi^2/d.o.f.=231.6/(171-4)=1.39$.
The fitted strength of the $S$-wave amplitude $\chi_S=-19.4 \pm2.5$~GeV$^{-1}$, the two other parameters of the scalar pion form factor being listed in  the caption of Fig.~\ref{fig:1}.
The fitted deviation from 1 of the strength of the $P$-wave amplitude
 is  $N_P=1.122\pm0.034$.

We obtain a good agreement with the experimental $m_{\pi^+ \pi^-}$ distributions as can be seen in Fig.~\ref{fig:2}.
We find a sizeable contribution of the $S$ wave just above the $\pi \pi$ threshold corresponding to the $f_0(600)$ contribution.
The $f_0(980)$ is not seen as a peak since the pion scalar form factor has a dip near 1~GeV as seen in  Fig.~\ref{fig:1}.
The signal for the  $f_0(1400)$ associated with the opening of the effective $(2\pi) (2\pi)$ channel (see Fig.~\ref{fig:1})  is tiny.
Looking at the distributions (see Ref.~\cite{ActaB42}) for the cosine of the helicity angle $ > 0$ and $< 0$,  one observes significant interference contributions, mainly between the $S$ and $P$ waves, under the dominant $\rho(770)^0$ peak.
%
\section{Conclusions and outlook}
\label{conclusions}

It is quite instructive to see that the quasi two-body QCD factorization allows i) to correlate heavy meson hadronic decay data with other meson-meson interaction data and ii) to give constraints between strong and weak interaction physics.
The determination of the strong interaction phases should help in the extraction of the weak angle phase $\gamma$ or $\phi_3 = \arg(-\lambda_u^*/\lambda_c^*)$.
Our $S$-wave amplitude  is proportional to $\Gamma _1^{n*}(s)$, its functional form could be used in Dalitz-plot analyses and  values can be sent upon request.
  
Here the data constrain strongly the non-strange pion scalar from factor and in particular its high-energy behavior.
It would be interesting if some semi-leptonic process could constraint the pion scalar form factor as in the case of the vector one.
On-shell non-strange and strange pion scalar form factors have been used to describe successfully the $\pi^+ \pi^-$ $S$-wave final state interaction in  $B \to K \pi^+\pi^-$~\cite{fkll,PRD74}. 
One should check if the  non-strange pion scalar form factor derived here together with the needed  strange component $\propto \left\langle [K \bar K]_{{S}} \ \vert s \bar s \vert 0  \right \rangle $
 would yield an equally succesfull description of the data. 

The $CP$ asymmetry 
for the $B$ to $\rho \pi$ decay 
is $A_{CP}[B^\pm \to \rho(770)^0\pi^{\pm},\ \rho(770)^0\rightarrow \pi^+\pi^-]=3.6\% \pm 0.2 \%$  which is  compatible with the experimental value given in Eq.~(\ref{ACP}). 
One expects  results with smaller errors from the Belle Collaboration, and hopefully, in the
near future, from LHCb and from the near term super $B$ factories.




\end{document}